\documentclass{article}

\PassOptionsToPackage{round}{natbib}
\usepackage[preprint]{neurips_2024}

\usepackage[utf8]{inputenc}
\usepackage[T1]{fontenc}

\usepackage{fvextra}  
\DefineVerbatimEnvironment{PromptBlock}{Verbatim}{
  breaklines=true,
  breakanywhere=true,
  breaksymbolleft={},    
  breaksymbolsepleft=0pt,
  fontsize=\small
}

\usepackage{bookmark}       
\usepackage{natbib}
\usepackage{url}            

\usepackage{booktabs}       
\usepackage{amsfonts}       
\usepackage{nicefrac}       
\usepackage{microtype}      
\usepackage[table,xcdraw]{xcolor}
\usepackage{csquotes}
\usepackage{svg}
\usepackage{tikz}
\usepackage{multirow}
\usepackage{tcolorbox}
\usepackage{listings}
\usepackage{adjustbox}
\usepackage{glossaries}
\usepackage{float}
\definecolor{backcolour}{rgb}{0.95,0.95,0.92}
\definecolor{customgreen}{rgb}{0.094, 0.502, 0.22}
\usepackage{graphicx}
\usepackage{enumitem}
\usepackage{makecell,array}
\usepackage{longtable}
\usepackage[capitalise]{cleveref}
\usepackage{hyperref}
\usepackage[toc,page]{appendix}

\crefname{appendix}{Appendix}{Appendices} 
\Crefname{appendix}{Appendix}{Appendices}

\lstdefinestyle{mystyle}{
    backgroundcolor=\color{backcolour},
    breaklines=true
}
\makeatletter
\AtBeginDocument{%
  \setcounter{secnumdepth}{4}%
  \setcounter{tocdepth}{4}%

  \renewcommand\paragraph{\@startsection{paragraph}{4}{\z@}%
    {1.5ex \@plus 0.5ex \@minus 0.2ex}
    {1.0ex}
    {\normalsize\bf}}%

  \let\subsubsubsection\paragraph

  \providecommand{\l@subsubsubsection}{\l@paragraph}%
}%
\makeatother

\title{A Methodology for Quantitative AI Risk Modeling}

\author{%
  Malcolm Murray \And Steve Barrett \And Henry Papadatos \And Otter Quarks
  \AND 
  Matt Smith \And Alejandro Tlaie Boria \And Chloé Touzet \And Siméon Campos
}

\begin{document}
\maketitle

\vspace{-0.7cm}
\begin{center}
    \small SaferAI
\end{center}
\vspace{0.7cm}

\begin{abstract}
Although general-purpose AI systems offer transformational opportunities in science and industry, they simultaneously raise critical concerns about safety, misuse, and potential loss of control. Despite these risks, methods for assessing and managing them remain underdeveloped. Effective risk management requires systematic modeling to characterize potential harms, as emphasized in frameworks such as the EU General-Purpose AI Code of Practice. This paper advances the risk modeling component of AI risk management by introducing a methodology that integrates scenario building with quantitative risk estimation, drawing on established approaches from other high-risk industries. Our methodology models risks through a six-step process: (1) defining risk scenarios, (2) decomposing them into quantifiable parameters, (3) quantifying baseline risk without AI models, (4) identifying key risk indicators such as benchmarks, (5) mapping these indicators to model parameters to estimate LLM uplift, and (6) aggregating individual parameters into risk estimates that enable concrete claims (e.g., X \% probability of >\$Y in annual cyber damages). We examine the choices that underlie our methodology throughout the article, with discussions of strengths, limitations, and implications for future research. Our methodology is designed to be applicable to key systemic AI risks, including cyber offense, biological weapon development, harmful manipulation, and loss-of-control, and is validated through extensive application in LLM-enabled cyber offense. Detailed empirical results and cyber-specific insights are presented in a companion paper.
\end{abstract}

\begin{figure}[t]
  \centering
     \includegraphics[draft=false,page=1,pagebox=mediabox, clip,width=\linewidth]{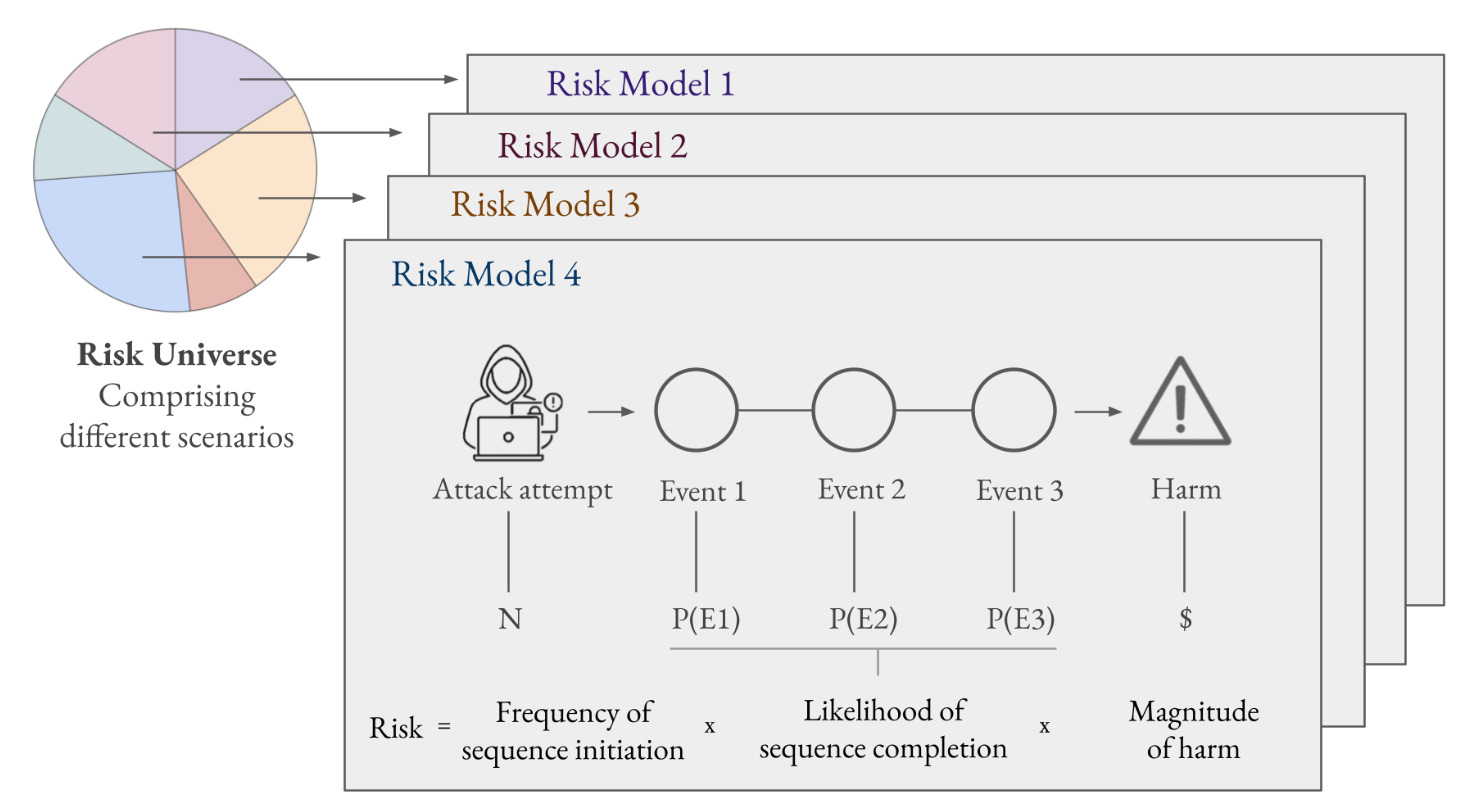}
 \caption{Our risk management methodology first decomposes the risk universe into distinct scenarios, then models each using three types of parameters: the frequency with which a specific sequence of events is initiated, the probability of the sequence taking place, and the harm that would arise as a result.}
 \label{Figure1}
\end{figure}

\section{Introduction}
Current risk management practices among frontier AI companies, as embodied in the frontier AI safety frameworks~\citep{metr_faisc}, focus on setting capabilities thresholds, assigning mitigations to each threshold, and conducting model evaluations to measure model capabilities~\cite{ukgov_international_ai_safety_report_2025, anthropic_rsp_updates, openai_preparedness_framework_v2_2025, deepmind_frontier_safety_framework_2_2025}. However, these frameworks do not measure thresholds for the actual risks, only for the model capabilities. These are only the source of risk. In addition, capability-based analyses often miss important factors related, for example, to threat actor behavior, target specificity, or the precise path to harm~\citep{lukosiute2025llm, ramanthresholds}. They also rely on imperfect measures of capabilities, measured by proxy through performance on benchmarks, which might be indicative of multiple capabilities~\citep{manifund_bayesian_llm_capabilities}.

Quantitative risk modeling, an essential component of risk management, addresses these gaps~\citep{Touzet2025}. It combines two aspects: (1) scenario building, where discrete, plausible risk pathways are chosen from the risk space and decomposed into discrete, measurable steps that link hazards (e.g., a model capability) to concrete real-world harms, and (2) risk quantification, where numerical estimates are assigned to each step. This results in outputs that are valuable for making specific claims. For example, “If a model scores above X on this benchmark, we can be 90\% confident that the expected annual damage will exceed \$1 billion”. This is highly useful for informing decisions related to AI deployment and is possible to compare with risk thresholds set by regulators or AI developers~\citep{koesslerthresholds, khlaafthresholds}.

The process of systematically building risk models, even before quantification, already provides substantial value. First, it creates a unified framework where diverse risk indicators (e.g. capability evaluations, benchmarks, red-teaming exercises, uplift studies, or incident reports) can be mapped to risk estimates. This framework ensures that evaluations are not designed in isolation, but rather contribute to a collective understanding of how real-world harms emerge. Second, risk models can also capture interactions between multiple capabilities. It might be the case that the risk emanating from a specific, single capability is below the level of concern, but that this changes when it is evaluated in conjunction with another capability or affordance. Third, risk models can guide evaluation priorities by identifying the largest sources of uncertainty and where additional evaluations would have the greatest impact. Finally, risk models can inform targeted mitigation strategies, for example, by revealing critical bottlenecks in risk pathways. If a specific LLM capability unlocks an entire risk pathway or a set of pathways, mitigation efforts can focus precisely on that capability.

Quantifying models provides further benefits. It enables stakeholders to compare AI risks with each other and put them in context by comparing them with better studied risks in other sectors. Second, quantification allows decision makers to create and enforce concrete risk thresholds. In aviation, for example, the FAA mandates that ``failure conditions which would prevent continued safe flight and landing'' must occur at frequencies below one per billion flight hours, equivalent to one catastrophic event every 114,155 aircraft years~\citep{faa_ac_25_1309_1a_1988}. Third, assessing a risk quantitatively instead of purely qualitatively allows a better identification of sources of disagreement and provides a clearer feedback loop to improve the risk assessment capabilities of developers. Qualitative risk modeling makes predictions using, for example, ``significantly'', which are much harder to falsify and, therefore, much less valuable to improve the risk model. For example, if risk models predict an increase by 10x for damages compared to the baseline, but no effect is visible on real-world year-over-year harm estimates, a review can easily be undertaken.

This paper presents a six-step methodology for creating quantitative AI risk models, which we demonstrate the practical applicability for by applying it to the domain of AI-enabled cyber offense risk. We do not include the full cyber risk models here; they are published in a companion paper focusing exclusively on the insights from the cyber risk models (Barrett et al., forthcoming). The remainder of this paper is organized as follows: \cref{sec:relatedwork} provides background on existing approaches and positions our contribution. \cref{sec:ProposedMethod} defines the quantitative modeling of risk and outlines our methodology in the six steps. \cref{sec:Discussion}, the discussion, presents use cases of risk models, and discusses limitations and areas for further research. \cref{sec:Conclusion} concludes the paper.
\section{Related Work}
\label{sec:relatedwork}
Our methodology is based on established risk modeling practices in other safety-critical domains. Quantitative risk modeling combines two components: scenario building and risk quantification. Scenario building is the foundational step in logically identifying the causal pathways that link a hazard to a potential harm. This is often done using structured techniques. Deductive, top-down methods like Fault Tree Analysis (FTA) start from a specific undesired outcome (e.g., a system failure) and work backward to identify all the combinations of root causes that could lead to it. Conversely, inductive, bottom-up methods like Event Tree Analysis (ETA) start from an initiating event (e.g., a component failure) and map out the branching sequences of possible consequences. Risk quantification then assigns numerical values to the likelihood and severity of the events within those scenarios using a combination of techniques. Expert elicitation is used to capture specialist knowledge and estimate probabilities where empirical data is lacking~\citep{apostolakis}. Monte Carlo simulations help propagate uncertainty through the model, generating a distribution of possible outcomes rather than a single point estimate~\citep{voseriskanalysis}. To formally manage uncertainty and update beliefs as new evidence emerges, Bayesian statistics are a standard tool. Specifically, to capture the complex interdependencies between events in a system and ensuring that the risk of the system as a whole is understood, methods like Bayesian networks (BNs) can be used to model probabilistic and causal relationships~\citep{bayesianfair}.

Quantitative risk modeling specifically applied to AI is still in its infancy. Yet, some approaches related to scenario building and risk quantification are emerging from academic and industry research. When it comes to structured scenario analysis, most efforts to date center on the development of safety cases, i.e., structured arguments, supported by evidence, making the case that an AI system is safe in a given context (see e.g.,~\citep{buhl2024safety, wasil2024affirmative, aisi_safety_cases_2024, clymer2024safety, goemans2024safety}). In industry practice, both Anthropic and Google DeepMind have begun to integrate safety cases into their research and governance frameworks~\citep{anthropic_safety_cases_2024, anthropic_rsp_v2_2_2025, deepmind2025}. This paper aims to add to this existing scholarship in various ways. In relation to safety cases, our methodology follows a chronological or causal logic rather than an argumentative logic and is designed to exhaustively map out all possible risk scenarios for an AI model as opposed to those related to a specific line of argument. The two approaches are highly complementary in that a robust safety argument for a high-risk system will likely reference or incorporate outputs from scenario building exercises. In a safety case arguing that a system is sufficiently safe to deploy, risk modeling outputs (i.e. risk scenarios and risk estimations) can be used as evidence to support a statement such as ``all key hazards have been identified and estimated''. A major difference between our methodology and attempts at quantifying safety cases~\citep{clymer2024safety, balesni2024towards} is that our underlying risk scenarios are more granular and comprehensive than argument-based safety cases.

Other research, for example, Convergence Analysis’s research program on “scenario planning”, focuses on tools for direct scenario development~\citep{convergence_scenario_research}. \citet{wisakanto2025adapting}, in their comprehensive Probabilistic Risk Assessment (PRA) for AI, suggest considering a model’s capabilities, domain knowledge, and affordances to systematically identify hazards, before modeling the risk pathways from these hazards, identifying causal sequences (via methods such as  FTA and ETA), and accounting for the effect of “propagation operators” (e.g., adversarial exploitation). This results in a systematic, but potentially overwhelming choice of scenarios. \citet{chin2025dimensional} also contributes to the scenario building scholarship on catastrophic AI risks such as  chemical, biological, radiological and nuclear (CBRN), cyber offense, and loss of control. His proposed methodology, which focuses on qualitative causal mapping, combines ``dimensional characterization'' to systematically analyze risks across seven key dimensions (including intent, competency, linearity, or reach) with ``risk pathway modeling'' to map out the step-by-step causal progressions from an initial hazard to a resulting harm. Compared to Chin's framework, our methodology is designed to facilitate quantification as a second step (notably by decomposing the scenarios into measurable steps).

Current attempts at risk quantification in the field of AI are also still nascent. Efforts are often limited to measuring model capabilities on specific benchmarks, which does not equate to an actual measure of the risk. \citet{murray2025mapping} propose a method to translate AI benchmark scores (derived from the cybersecurity benchmark Cybench~\citep{zhang2024cybench}) into risk estimates using the IDEA protocol~\citep{hemming2018practical} for structured expert elicitation and creating a direct mapping from benchmark performance to real-world risk. Beyond capability-based approaches, more sophisticated approaches are emerging, such as the analysis presented by~\citet{rodriguez2025framework} of how AI helps cyber attackers. They start by identifying representative attack scenarios, using Lockheed Martin’s Cyber Kill Chain and drawing on more than 12,000 real-world AI-powered cyber incidents. They use expert elicitation to identify “bottlenecks”, defined as any step requiring at least 10\% of the total estimated resources for the attack. Finally, they measure how much AI reduces the cost of executing these bottlenecks. In this way, they are able to identify the precise steps where AI empowers cyber attackers the most. Similarly, our methodology also considers a “basket” of potential risk models, but we place a greater emphasis on estimating the likelihood of each step and a smaller emphasis on estimating cost. We believe that likelihood can similarly be used to identify bottlenecks and might be the more universal of the two measures across risk domains.

\citet{righettibio} and~\citet{halsteadcyber} have produced quantitative estimates for hypothetical benchmark results and uplift studies of catastrophic AI-enabled risks. Under constraints such as "if AI systems were to increase by 10 percentage points the number of STEM undergraduates able to synthesize pathogens as complex as influenza", they estimate how much the probability would increase for a risk scenario such as an epidemic caused by a lone-wolf terrorist attack. 

Others have explored quantification within safety cases, proposing methods for assigning probabilities to claims and aggregating them to produce an overall confidence estimate, sometimes expressing the claims themselves in quantitative terms~\citep{clymer2024safety, balesni2024towards, clymer2025example}. \citet{barrett2025assessing} propose a safety case framework to combine evidence on threats as well as the effectiveness of mitigations (in their example, API-based safeguards) to produce an overall quantitative estimate of risk. One challenge associated with quantification within safety cases is that achieving high confidence in a top-level claim requires extremely high confidence in every subclaim, and simplistic aggregation methods often rely on problematic assumptions of independence among arguments~\citep{balesni2024towards, barrett2025assessing}. We note that the difficulty in achieving high confidence in top-level claims identified in other papers also partly applies to our methodology. However, we believe our methods for quantification of both probabilities and quantities with confidence intervals, and Monte Carlo propagation of estimates, enables us to nevertheless make top-level claims with high confidence.

\citet{wisakanto2025adapting} propose a semi-quantitative approach to risk estimation using coarse-grained bands (structured to span orders of magnitude) rather than precise probabilities. They use a ten-level risk matrix, combining information on harm severity levels (from marginal to globally catastrophic) with information on likelihood levels characterizing the probability of occurrence with defined odds bands. For likelihood estimation, they suggest applying the following formula: "P(harmful scenario) = P(capability exists) × P(capability misused | exists) × P(harm occurs | misused)", whereas our approach looks at probabilities for each step of an event sequence. 

\section{Proposed Methodology}
\label{sec:ProposedMethod}
Our methodology is based on best practices in the literature on risk management and refined through discussions with experts in risk management, AI, and forecasting. It is developed through the iterative creation of risk models specifically in the domain of cybersecurity. Our aim is for the methodology to be applicable to all of the risk domains referred to in the EU AI Act’s Code of Practice~\citep{ec_contents_code_gpai} as systemic (chemical, biological, radiological and nuclear; cyber offense, loss of control; and harmful manipulation), and we have taken into account the characteristics of all these risk areas. However, we do not claim that it is applicable to more diffuse risks that rely on cascading effects over time, such as labor market disruption. Most of the examples below are taken from the cyber-offense setting, as that is where we have done the most extensive testing. We start with cyber as a domain as it is a field where AI is already increasing the level of risk~\citep{anthropiccyberreport} and it lends itself to quantifying risk\citep{fairframework}. In the companion paper~\citep{Barrettb2025}, we provide a detailed study of our applications of the method developed here in this setting, alongside empirical validation of our approach, and cybersecurity-specific insights derived from our model. There is currently a paucity of data on the realized risks  that can be directly linked to the use of AI systems, which makes it challenging to empirically validate risk models (an exception being biological weapons, where the lack of attacks in recent years sets an upper bound). This current work seeks to establish a framework for proactive risk modeling despite these constraints. Although we expect many of the principles developed here to generalize across risk domains, specific adaptations may be needed in other domains. More extensive testing of the cross-domain validity will become possible as more data becomes available.

As seen in~\cref{Figure1}, our methodology is rooted in the definition of risk as harm arising from a certain sequence of events. Risk can then be calculated as the product of three terms, (1) the frequency of occurrence of event sequence-triggering events, (2) the probability that the entire event sequence will occur and (3) the magnitude of the harm arising. It combines a top-down approach of choosing risk scenarios with a bottom-up approach linking empirical data to specific parameters. For a given AI risk domain, it can prove useful to break down the complete risk universe associated with that given risk domain into a number of different risk scenarios, where a separate risk model is developed for each scenario.

The key principles underlying the methodology are:

\textbf{Decomposing risks into the components of the risk equation}. In order to capture the different ways in which LLMs affect risk, we estimate separately the frequency of initiation of the event chain, the probability of success of the event chain and the harm that results from a successful completion of the event chain. This is a common practice in risk management across several domains~\citep{iso_iec_guide_51_2014, kaplan1981quantitative}.

\textbf{Outlining the sequence of events that occur to turn a hazard into harm}. In order to accurately capture all the events that need to occur for an initial hazard (such as a dangerous model capability) to result in harm (e.g., monetary or harm to persons), we define a sequence of events. This is a common risk management practice~\citep{apostolakis1990concept, sra_fundamental_principles}. Typical examples of how this can be done include Fault Tree Analysis and Event Tree Analysis~\citep{iec_61025_2006_fta, nrc_nureg_0492_1981}).

\textbf{Leveraging expert elicitation through a (modified) Delphi process.} Given the paucity of historical data, we make use of expert elicitation to estimate parameters. This is a common practice in e.g., the nuclear power industry~\citep{apostolakis1990concept, xing_morrow_2016_expert_elicitation}. We follow the IDEA protocol~\citep{hemming2018practical}, which is a modified Delphi approach~\citep{hsu2007delphi, cooke2004expert}, and has been used across domains such as environmental risk assessment and political science.

\textbf{Probabilistic estimation of parameters.} Given the high level of uncertainty about the trajectory of AI and its impacts, we estimate each parameter as a three-point estimate (i.e. mode and confidence interval) rather than a point estimate. This is common in Bayesian analysis~\citep{apostolakis1990concept, pate1996uncertainties}.

\textbf{Statistical aggregation of parameters.} To capture the large natural variation of risk outcomes, we aggregate the estimates through a statistical method, involving fitting the parameters to distinct distributions and running Monte Carlo simulations. This is commonly used in risk assessment in financial services as well as in the nuclear power industry~\citep{vose2008risk3e, de2019deterministic}.

\textbf{Using LLMs as estimators.} Given the lack of experts in the cross-section of AI and specific risk domains, we explore the use of LLMs as estimators (after validating their results compared to human experts). This is a nascent approach that has only recently become salient~\citep{halawi2024approaching}.

\textbf{Create a mapping between risk indicators and risk.} In order to make the models more future-proof and scalable, we derive a relationship between KRIs and risk estimates. This allows for a simpler estimation process in future iterations (see e.g.~\citep{murray2025mapping}).

Our methodology consists of six closely interlinked steps:
\begin{enumerate}
    \item \textbf{Defining risk scenarios to model.} We systematically decompose the risk universe into a set of representative scenarios to build models for., and we build risk models for each representative scenario. 
    \item \textbf{Constructing risk models.} Risk is modeled as a combination of 4 factors. First, the number of threat actors conducting this type of attack. Second, the number of attack attempts per actor per year. Third, the set of tactics required in the attack and their associated probability of successful application. Fourth, the damage resulting from each successful attack. 
    \item \textbf{Quantifying "baseline risk".} We establish estimates for the risk of “baseline” threat actor capabilities (negligible or no use of AI) as a reference point for uplift. 
    \item \textbf{Determining key risk indicators for AI “uplift”.} We establish which forms of evidence (KRIs - Key Risk Indicators), such as benchmark performance, risk model factors can be conditioned on. 
    \item \textbf{Estimating AI uplift.} We conduct expert elicitation to build a quantitative mapping between the KRIs and the factors in the risk model and use these to generate uplift estimates. 
    \item \textbf{Propagating individual estimates to aggregate estimates.} Distributions over risk factor parameters are fitted to expert epistemic uncertainty. We propagate and aggregate - across experts and risk factors - samples from these distributions through Monte Carlo simulation to estimate a distribution over the overall risk.
\end{enumerate}
In the below, we go into details regarding each of these steps.
\subsection{Step 1: Defining Risk Scenarios to Model}
\label{step1}
In the first step, we determine which risk domain to model and which scenarios to model in the domain. The choice of risk domain can guide decisions about what type of harm to use and in what units to measure it. For our experiments in applying the methodology to cyber offense, the type of harm is economic damage, captured in US dollars.

To enable decision makers to make go/no-go decisions for the development and deployment of AI models, risk models must be representative of a sufficiently large part of the risk space. However, the risk space is vast and not all scenarios can or need to be modeled. To achieve this, we decompose the risk space using three taxonomized aspects - actor, target, and vector, analyze the ways they can be combined, and select the most relevant combinations.  For each of these aspects, we believe it is best to use established taxonomies where available to enable standardization and replication by others.

\textbf{Actors.} The threat actor leverages a hazard and turns it into a harmful event. In the field of cybersecurity, a useful taxonomy to use is RAND’s classification of offensive cybersecurity operations along a spectrum of OC1 to OC5, from amateur attempts by hobbyist hackers to top priority operations by the most cyber-capable nation states~\citep{nevo2024securingweights}.

\textbf{Targets.} The target is the entity that suffers harm. A useful taxonomy here is the list of critical national infrastructure sectors from CISA in the US and NPSA in the UK~\citep{cisa_critical_infrastructure_sectors, npsa_cni_page}. These include sectors often targeted by malicious users, such as financial services, healthcare, transportation systems, and defense.

\textbf{Vectors.} The vector is the type of attack. In the case of cybersecurity, a good starting point is~\citet{rodriguez2025framework}, who analyzed real-world instances of AI use attempts in cyber attacks with a large data set of incidents.

Combining these actors, targets and vectors could result in a large number of possible combinations (560 using the examples above). We therefore apply a set of principles to choose a smaller set of the most salient combinations. First, we look at historical data and plausibility to prune the number of scenarios.

\textbf{Historical data.} We pick scenarios that are representative of the actors and where historical data, where available, shows the scenario has been prevalent. However, we also include a few novel aspects to achieve greater diversity in the scenarios.

\textbf{Plausibility.} We remove uncommon or unrealistic combinations. In the case of cyber attacks for example, OC1 actors will typically not attack well-guarded defense targets.

Second, to choose the most salient risk models, we ensure we include risk models that capture the ways in which LLMs help the actors. \citet{rodriguez2025framework} outline three ways in which AI models help malicious actors: Capability Uplift, Throughput Uplift and Novel Risks.

\textbf{Capabilities.} We look at which LLM capabilities are expected to be the most influential on the level of risk and pick risk scenarios accordingly.

\textbf{Throughput.} While specific capabilities of LLMs can be said to increase the “quality” of the attack, LLMs also increase the “quantity”. LLMs enable malicious actors to conduct certain activities at much greater scope and scale.

\textbf{Novelty.} A further angle to use is what unique and novel risk scenarios were previously impossible that now may become possible using LLMs. For example, in the case of cyber attacks, LLMs may enable overcoming previously-sufficient defense mechanisms by scaling up the cadence and breadth of the attack. This could mean more attacks on e.g., electrical grids or the banking system.

By applying these principles, we derive a suitable set of risk scenarios for which to build risk models. Throughout the process, we validate the choices of scenarios with cyber experts.
\subsection{Step 2: Building Risk Scenarios}
\label{step2}
\begin{figure}[t]
  \centering
     \includegraphics[draft=false,page=1,pagebox=cropbox,trim=0 1cm 0 0, clip,width=\linewidth]{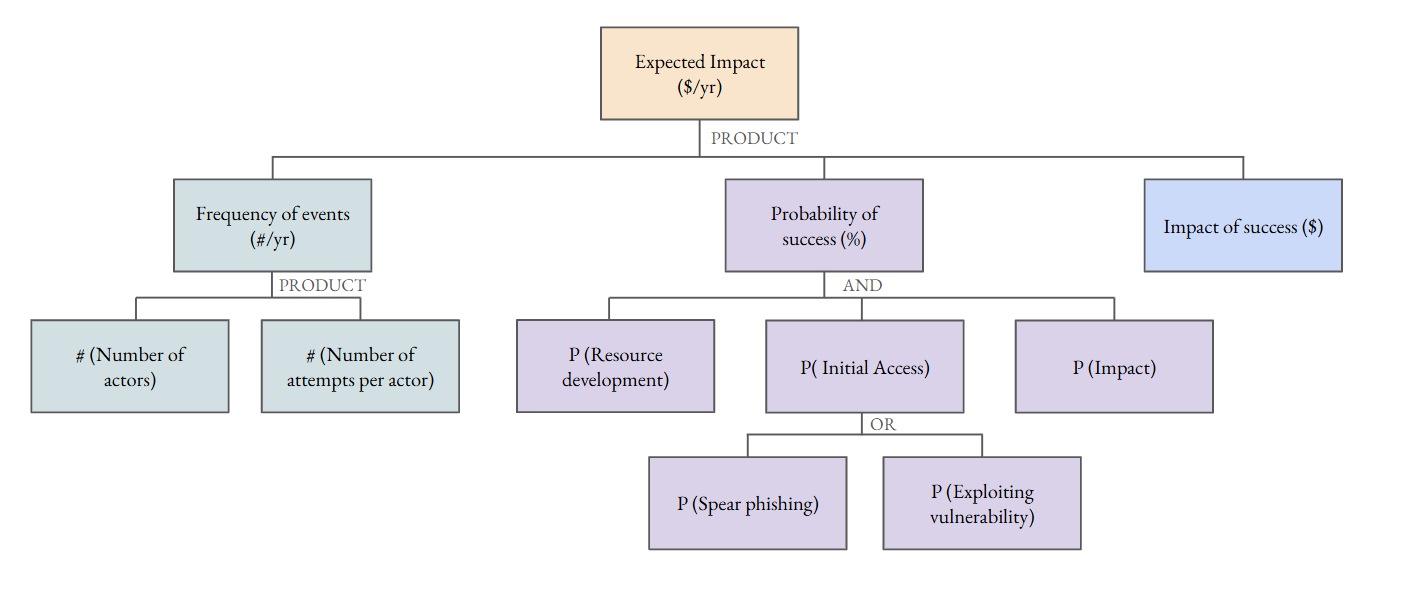}
 \caption{An illustrative risk model decomposed into its constituent parameters.}
 \label{Figure2}
\end{figure}

Having defined our risk scenarios of interest by systematically filtering the universe of risks for a given domain, we proceed to build out detailed risk scenarios. This step of the process takes the high-level risk scenarios that we have identified previously, and builds them out into complete sets of parameters and steps that fully capture the risk and sequence of events leading to harm (see~\cref{Figure2}). 

Since the subsequent step of the methodology involves quantifying the risk, this model-building step needs to break down scenarios to an appropriate level of granularity. This should balance measurability or ``estimatability'' of parameters with fidelity to real-world events.

There are several benefits of increasing the granularity of the model. Breaking parameters into smaller components makes each one more self-contained and straightforward to estimate. A single high-level parameter often requires experts to internalize multiple factors, reducing accuracy. There are also benefits with respect to the outputs that a granular model can produce; more detailed models reveal the mechanisms driving the risk, and highlight non-obvious interactions between components. Granular models also enable better sensitivity analysis, help prioritize mitigation by identifying critical steps, and show where further research would most reduce uncertainty. Since benchmarks typically evaluate narrow capabilities, granular parameters can map directly to individual benchmarks.

Increasing granularity, however, also trades off model accuracy. Experts can generally estimate high-level parameters (e.g., "Will a nation-state actor succeed at getting initial access?") by drawing on implicit knowledge of the outcome of the step. Breaking this down forces explicit modeling of specific pathways, which will inevitably miss some real-world possibilities that implicit expert judgment might have accounted for. When experts estimate narrow sub-components, they may also each envision different mechanisms. These varying assumptions become problematic when sub-estimates are combined, potentially yielding results that do not represent any coherent real-world scenario. Higher-level estimation avoids this as all experts estimate the same overall outcome, regardless of the specific paths they imagine. Increasing granularity also increases the cost of the full modeling methodology, which can rapidly become prohibitive.

In an attempt to arrive at an appropriate level of granularity, we apply a principled approach to breaking down each risk model. We first decompose the risk into its constituent parts: the frequency of occurrence of the initiating events, the likelihood of success of the sequence of events, and the impact of the event in the case that it does succeed. Where appropriate, we further break down each of these high-level parameters into sub-components, in an effort to increase their estimatability. For misuse risk models for instance, we believe it is useful to break down the frequency of occurrence into two parameters - the number of actors who might attempt to cause such an event, and the number of attempts each actor may perform within a given time frame. This allows us to account for two very different considerations regarding the impact of LLMs - larger incentives to attack as well as higher speed of attacks. Estimating these separately increases accuracy without introducing much additional complexity.

We generally break down the probability of success parameter to a fairly high level of granularity, because it is generally the most complex parameter with the most information. In addition, it provides the greatest insights into the relative effects of mitigations and the need for new benchmarks measuring specific steps. To break down the probability parameter, we follow a process akin to Fault Tree Analysis (FTA), whereby we start with the occurrence of an event causing an impact, and break this event down into its necessary or sufficient components.  To avoid over-complication of the model, we follow a number of heuristics: 
\begin{itemize}
    \item We generally repeat the process of splitting steps into sub-components until we deem each step to be estimable by an expert, and that we have captured all of the different AI-specific mechanisms that a step can contain.
    \item We avoid the introduction of steps that introduce dependencies (where different outcomes at a node would lead to different attack paths). 
    \item We avoid introducing nodes that are not necessary to the success of the overall scenario (“nice to haves”).
    \item We avoid nodes whose failure modes are redundant with earlier steps to avoid double counting a failure mechanism. 
    \item We avoid splitting nodes when the act of choosing which specific sub-components to model introduces more error than letting experts implicitly consider all possible pathways in their high-level estimate. This is typically the case when there is data available describing the output of the node, but not the implicit mechanisms leading to the output.
\end{itemize}
\subsection{Step 3: Quantifying Baseline Risk in Risk Scenarios}
\label{step3}
We then turn to quantifying the risk scenarios. We first quantify “baseline risk”, i.e., the risk in the absence of the use of LLMs. This initial estimate is used to enable calculating “marginal risk”, i.e., how much risk is added when the use of LLMs are included in the scenario (note that this applies somewhat differently in the case of loss of control scenarios where there is no "baseline"). We estimate each parameter in the risk scenario for the non-LLM scenario. In doing so, we rely heavily on “base rates” (the historic rate or frequency with which an event has occurred) and other statistical information to inform the estimates. We also draw heavily on historic case studies to ensure the models close track real-world events. We notably rely heavily on historical accounts and incident reports. We also make heavy use of domain expert feedback during this process, ensuring each scenario is reviewed at least once fully by an expert with domain-specific experience relevant to the risk in question, using their feedback to iterate and refine the model. To validate the methodology for our expert reviewers, we ran an estimate falsification experiment on one of our cybersecurity expert reviewers. For three of the twelve probabilities that he had to estimate, we falsified the probability estimate along with its rationale, before sending them to him. For these three, the expert reacted to the falsification and argued that the estimates were incorrect. This demonstrates that, at least in the cyber realm, the process works as planned. The full logs of this experiment are included in~\cref{app:C}. 

\subsection{Step 4: Identifying and Pre-processing Key Risk Indicators (KRIs) for LLM Uplift}
\label{step4}
Next, we turn to estimating LLM uplift. Equipped with parameterized model scenarios capturing realistic dynamics of risk events, we must first establish which forms of evidence can be used to estimate the LLM uplift. We lack direct evidence that could be used to estimate values, such as the real-world rate of occurrence of a particular risk factor. We therefore make use of Key Risk Indicators (KRIs), quantifiable measurements of AI system behavior that can serve as indirect evidence for model parameters. KRIs provides the inputs that ground our risk models in the real-world behavior of AI systems, enabling more accurate risk estimates. We build a quantitative mapping between LLM capabilities and the values of the parameters in the risk model.
\subsubsection{Selection of KRIs}
Currently, benchmarks are the primary form of KRIs that are suitable for risk modeling in the context of AI systems. A benchmark is a standardized set of tasks used for the quantitative evaluation of capabilities across a range of AI systems. Generally, for a given AI system, a wide variety of benchmark data can be collected in order to capture the capabilities of the model, which are then used to estimate parameters in the risk model. Other possible forms of KRI data include:
\begin{itemize}
    \item \textbf{Red-teaming:} A controlled process where experts attempt to elicit specific harmful behaviors from an AI system to identify vulnerabilities.
     \item \textbf{Uplift Studies:} An evaluation of human performance on a task with and without AI assistance, to quantify the capability gain an AI system provides, particularly for misuse risks.
      \item \textbf{Incident Reports:} Documentation of real-world events where an AI system contributed to a risk event.
\end{itemize}
Red-teaming and uplift studies are the most comprehensive options, but these are resource-intensive to design and execute. Their evidence is also model-specific and quickly becomes obsolete due to rapid AI development cycles. Benchmarks are designed to support automated and low-cost evaluation across AI systems, and provide improved longevity. AI-specific incident reports are currently too scarce to support robust quantitative analysis, and existing incident reports are not necessarily informative of the behaviors of new models, so are less future-proof. \\

Consequently, this study focuses on the use of benchmarks, which can be applied to a wide range of AI systems, and are inexpensive and widely available. We note, however, that in principle, as the evaluations ecosystem matures and more incident data becomes available, these KRIs should be incorporated into risk models in a similar way to benchmarks.

We select benchmarks to serve as KRIs based on three primary desiderata:
\begin{itemize}
    \item \textbf{Unsaturated:} A benchmark must effectively differentiate between AI systems. Benchmark saturation occurs when models achieve near-perfect scores, rendering the benchmark useless for measuring differences in capabilities between models. This can result from generalized capability growth or from models training directly on benchmark data. Therefore, we only select benchmarks where top-performing models score significantly below the maximum.
    \item \textbf{Community Validation:} The benchmark must be accepted as a meaningful measure by the research community. We gauge this through proxies like high citation counts and inclusion on influential leaderboards. These indicate that a benchmark is well-formulated and accurately measures its stated capability.
    \item \textbf{Risk Relevance:} The tasks in the benchmark must be demonstrably relevant to the risk model's parameters. We select benchmarks whose tasks correspond directly to the behaviors described in our risk scenarios. This relevance is independently verified by domain experts prior to parameter estimation.
\end{itemize}

Together, these desiderata ensure our KRI selection is robust, validated, and informative. In addition, in the present study, we supplement these with three simplifying constraints. These allow for simpler risk models and easier estimation procedures, but we note that these could be relaxed in future work: 
\begin{itemize}
    \item \textbf{Static Scoring:} To ensure comparability across different models and times, we exclude benchmarks that use a dynamic scoring model (i.e., another AI system) to evaluate performance. Such methods are brittle, as changes to the AI system used for scoring invalidate historical results.
    \item \textbf{Realistic, Self-contained Tasks:} In order to facilitate estimation, we select benchmarks with tasks that assess the AI system in a realistic setting. We avoid benchmarks composed of a vast number of small tasks (e.g., multiple-choice questions). In order to inform a risk parameter, many questions would have to be combined, introducing unnecessary estimation complexity. We prefer benchmarks with a smaller number of substantive, independent tasks.
    \item \textbf{Rankable Difficulty:} To streamline the expert elicitation process, we also currently require that a benchmark's tasks can be ordered by difficulty (defined by the success rate across a representative model sample, assuming a monotonic relationship where success on a difficult task implies success on all easier ones). This allows for simpler elicitation models that do not need to account for complex task interactions.
\end{itemize}
\subsubsection{Pre-Processing of KRIs}
Once a set of KRIs have been selected, we conduct several pre-processing steps which establish the details of how benchmark scores will be used in later steps to estimate risk model parameters.

\subsubsubsection{Automatic Difficulty Ranking}

To facilitate expert elicitation (\cref{step5}), for each benchmark, tasks are ordered in accordance with their difficulty. Some benchmarks include a natural difficulty metric (e.g., the "First Solve Time" of capture-the-flag tasks in the Cybench test suite~\citep{zhang2024cybench}), but most do not, necessitating an estimation of this ordering.

In order to estimate task difficulty in a scalable manner without requiring extensive expert knowledge, we propose an automated procedure by which tasks can be ranked according to their difficulty. To do so, we develop a procedure for automated ranking of task difficulties using an LLM:
\begin{enumerate}
    \item The LLM assigns a difficulty score to each task in isolation. The scoring is anchored to a domain-specific scale (e.g., 0-100) using provided examples of low- and high-difficulty tasks.
    \item The LLM is presented with the full set of task descriptions and identifies the easiest one. This task is ranked lowest and removed from the set. The process repeats until all tasks are ranked.
    \item The inverse of the above; the LLM repeatedly identifies the most difficult task from the remaining set to build a ranking from hardest to easiest.
\end{enumerate}

These rankings are aggregated using the Borda Count algorithm, to produce a final difficulty ordering. The difficulty estimation procedure is further validated by domain experts by providing them with the task descriptions, and asking them to rank the overall difficulties of the tasks manually. Ordering metrics such as Kendall's W correlation~\citep{kendall1939problem} is used to compare expert task ordering with the estimated ordering, and we see high correlation, indicating that an accurate ranking has been achieved.

\subsubsubsection{Assignment of KRIs to Parameters}

In order to make the best use of existing information in estimating specific parameters across the models, we pick specific KRIs for each parameter. To do this, we consider which benchmark contains tasks which are most relevant to the particular step and the parameter. We define relevance as the similarity between the benchmark tasks and the risk scenario parameter and use the benchmark with the greatest number of tasks relevant to the risk step to estimate the parameter of the risk model. In many cases, this overlap will naturally be imperfect - an ideal benchmark would contain tasks leveraging all the skills needed to execute this action.

\subsubsubsection{Sub-sampling Tasks}
As some benchmarks have a large number of tasks, we pick the most relevant among them, in order to reduce the amount of expert elicitation needed. The sampling strategy is carefully designed to create a representative range of task difficulty in the benchmark and minimize the amount of data loss in the elicitation process. To accomplish this, we first sample tasks at fixed intervals along the difficulty-ranked list of tasks, ensuring an evenly distributed and maximally broad range of task difficulties. For each selected task index n in the difficulty ranking, we evaluate the local neighborhood of tasks in the original list (tasks from position n-1 to n+1). If replacing a task in the subsample with one of its neighbors improves the diversity of the overall sample (in terms of how many distinct concepts are covered by the benchmark tasks), a swap is made. For example, in the cyber attack misuse setting, a swap was made wherever a task had significant overlap with the rest of the subsample, and swapping would add a task to our subset that included new skills and attack vectors not present in the rest of the subsample. This process is repeated, expanding the local neighborhood of the original sample until the subset of tasks is sufficiently diverse.

\subsection{Step 5: Estimating LLM uplift in risk scenarios}
\label{step5}
Next, we apply the evidence to estimate LLM uplift through expert elicitation. We leverage human experts in a Delphi study as well as "LLM experts" in a simulated Delphi study. 

\subsubsection{Expert Elicitation}
\begin{figure}
  \centering
        \includegraphics[draft=false,page=1,pagebox=cropbox,trim=0 1cm 0 0, clip, width=\linewidth]{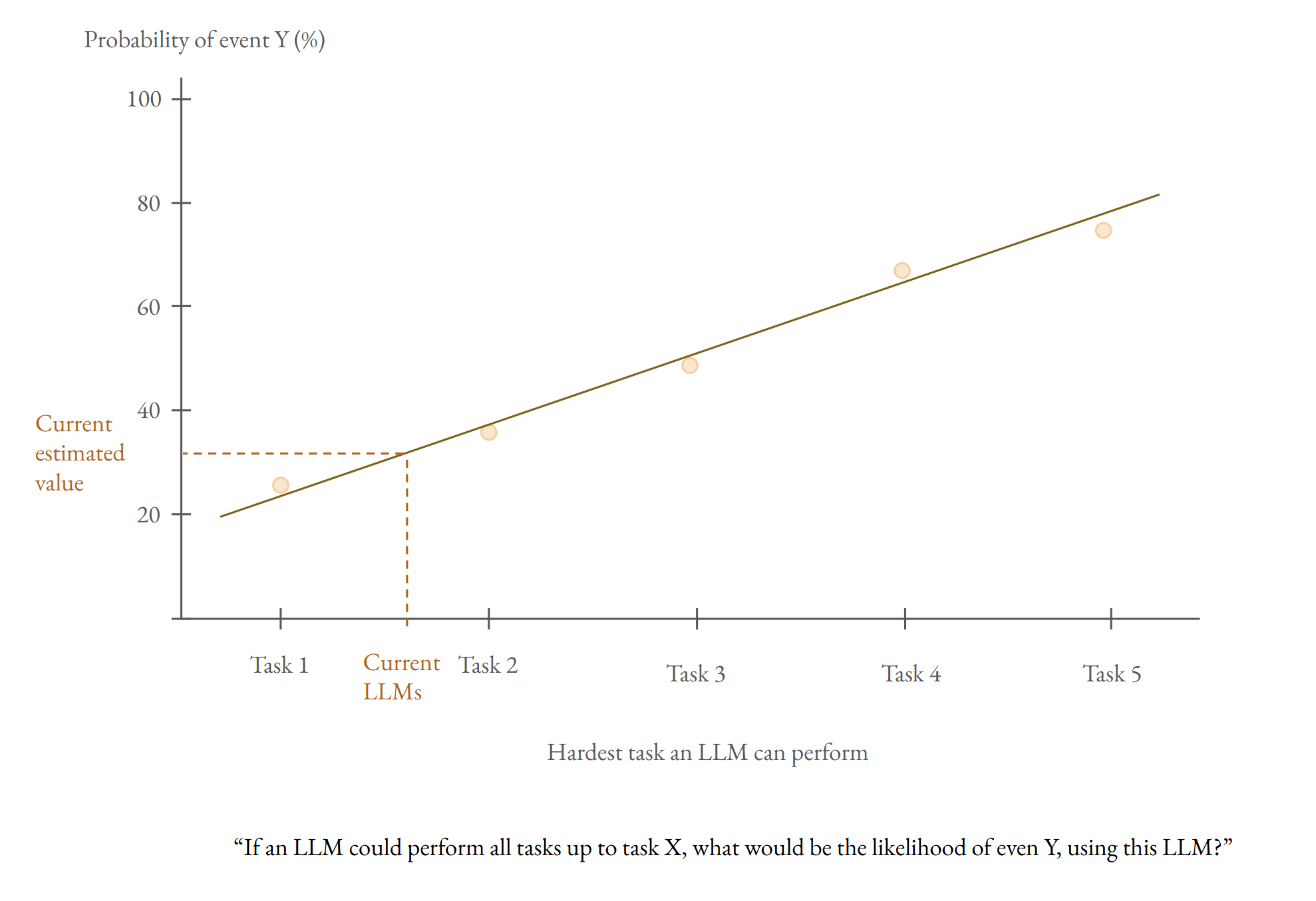}
 \caption{An illustrative mapping of benchmark performance to risk parameter value, constructed by elicitation with the question “If an LLM could perform all tasks up to task X, what would be the likelihood of event Y, where this LLM is used?” and interpolated.}
 \label{Figure3}
\end{figure}

Having assigned an indicator to each parameter in the risk model, we seek to build a quantitative mapping between LLM capabilities and the values of the parameters in the risk model. The purpose of this stage is to produce a model such that for an LLM with any given capability, the level of risk it produces can immediately be quantified. A direct mapping where risk could be fully explained as a function of LLM capabilities would be desirable. However, it is not feasible since the dynamics linking capabilities to risk are highly complex. Similar to other high-risk industries such as the nuclear power industry~\citep{xing_morrow_2016_expert_elicitation}, we therefore rely on expert elicitation as an intermediate step, to build a discretized mapping of different capability levels to different values of each parameter. This can later be inter- and extrapolated to construct an explicit function (see~\cref{Figure3} for an illustrative example). 

In order to elicit the mapping of capabilities to risk, we identify experts with domain expertise in both AI and the risk domain of interest and ask them to estimate, for each benchmark task, \textbf{if this is the most advanced task that the LLM can solve and it is available in the scenario, what is the corresponding value of the parameter?} For cyber, for example, we may ask “If X task is the hardest coding task that a theoretical LLM can solve, what is the likelihood that a cyber crime group with full access to this LLM could develop this type of malware?” By “hardest benchmark task an LLM can perform”, we mean that the LLM can solve this task more than 9 times out of 10, as well as tasks at a lower difficulty level, while it will most often fail or make mistakes on any harder tasks. We also encourage the experts to infer what broader capabilities an LLM would possess if it could solve all tasks up to and including this difficulty level, rather than focusing on the specific skills required to solve a given task. At this stage, we also provide the experts with as much contextual information as possible that we gathered during the scenario building (base rates, incident reports, historically similar cases, etc.) For each parameter, we aim for approximately five different capability levels (correspondingly, five benchmark tasks of increasing complexity).

We apply this process to each type of parameter in the risk model, and encourage the experts to consider the relevant effects of LLMs on each parameter:
\begin{itemize}
    \item For the frequency parameters, experts must consider the effects of LLMs both on scale (enabling more actors or attempts) and access (enticing less sophisticated malicious actors).
    \item For probability estimates, experts must consider what broader capabilities can be inferred from the capability indicator (the benchmark task), which of these help overcome existing defenses in this step, and which defensive measures remain effective regardless of increased LLM capabilities.
    \item For impact parameters, experts must consider the cumulative effect of LLM assistance throughout the event on any harm-specific elements, such as the size of ransom in the case of cyber.
\end{itemize}
For elicitation, we make use of the IDEA (Investigate, Discuss, Estimate, Aggregate) protocol, a modified Delphi process~\citep{hemming2018practical}. The process has four steps:
\begin{enumerate}
    \item Experts first make estimates individually
    \item Then all the experts collectively discuss their first estimates
    \item Then experts have the opportunity to update their initial estimates based on the discussion
    \item Finally, we aggregate the results from the second estimates
\end{enumerate}
For each parameter, we also utilize the four-step elicitation protocol outlined in IDEA, which is used to reduce overconfidence in expert elicitation. For each combination of parameter and capability level, we ask experts to estimate their best guess, the lowest and highest plausible values, and their confidence that the real value lies between these bounds. In this manner, each estimate captures the best guess of the expert, and a confidence interval determined by the expert. These confidence intervals reflect the epistemic uncertainty of the expert within the bounds of the model, not the aleatory variance of a given parameter (see further details in~\cref{app:D}).

As domain experts may lack background experience in reasoning in probabilistic estimation, we also provide them with best practices for forecasting, well-defined phrasings of the questions to avoid confusion, simple calibration exercises and superforecasting experts who probe the experts’ rationales in the discussion phase.  
We perform this process for all the parameters in a model, and we conserve the same group of experts to perform the entire elicitation process, as many of the parameter estimations benefit from having estimated the other parameters in the same model. As models often contain tens of different parameters to be estimated, each of which requires estimates for five different levels of capability (benchmark tasks), we perform the two estimation rounds asynchronously and over a longer period of time. This helps prevent expert fatigue and avoids the difficulty of gathering large numbers of domain experts for a set time period. In order to further alleviate the resources required, we pick a small set of parameters whose uplift do not need to be estimated from a function of the LLM’s capabilities. For example, if the variance in a parameter is fixed and low (e.g., when estimating a frequency of initiating events that is bounded by considerations other than the LLM capability, or estimating a probability which has a very high baseline already before LLM improvement), estimating their uplift adds little value. In the case of cyber, this means not measuring the uplift to probabilities that have a baseline value above 85\%. The process remains resource-intensive, however, requiring up to a month of part-time commitment from experts with a rare combination of expertise to complete just one risk model.
\subsubsection{LLM Estimation}
The protocol outlined above yields high-quality estimates, but is resource-intensive and difficult to scale across many parameters, capability levels, and domains. We therefore also explore an LLM-simulated elicitation that mirrors the same structure (analysis → estimation → aggregation), while allowing rapid, repeatable estimation runs. The goal is to generate estimates along with structured rationales explaining the reasoning for the estimates.

Task descriptions provided in benchmarks are often heavily summarized, sometimes to a single sentence or paragraph per task. As we use these benchmark tasks as a proxy for the capability ceiling of a theoretical LLM, we must ensure that the LLM we use to perform the quantitative assessments (hereafter LLM-estimator) has sufficient understanding of these tasks. We therefore employ a two-stage prompting approach that I) analyzes the benchmark task and the technical capabilities that an LLM capable of completing such a task would possess, and II) produces calibrated probability estimates.

We begin by prompting the LLM-estimator to perform a comprehensive analysis of the benchmark task, requiring it to detail the technical capabilities needed for execution, assess how an LLM with sufficient capability to solve the task would impact real-world scenarios, and evaluate the practical assistance such an LLM could provide. This analysis incorporates FST (First Solve Time) metrics from CTF (Capture the Flag) competitions as concrete difficulty indicators.

We then feed the analysis output into a structured estimation prompt that guides the LLM-estimator through a three-phase reasoning process, inspired by the work of~\citet{halawi2024approaching}. The prompts can be found in~\cref{app:B}. The LLM-estimator must:
\begin{enumerate}
    \item Leverage all available information to establish reasonable probability ranges, considering both the lowest and highest plausible success rates. 
    \item Critically evaluate its initial assessment for potential over- or under-confidence.
    \item Examine the practical implications of the suggested probability improvements and test how reasonable the estimate is against real-world expectations. 
\end{enumerate}
The LLM-estimator outputs its estimate supported by a summarized rationale that explains the reasoning behind the quantitative assessment and the factors that most significantly influence the probability estimates. Using the LLM’s system prompt, we instantiate the LLM-estimator with varied profiles of experts in cybersecurity. \citet{barrett2025assessing} identified that including different expert profiles allows capturing different aspects of a task, and improves predictions. Their study used LLMs prompted with 50 expert profiles with high-level role descriptions (e.g., Economist - Macroeconomics). This approach achieves variance through quantity and breadth of disciplines. Since we aim to ultimately perform many estimations, using 50 expert profiles per estimation would result in excessive computational time and cost. For this proof of concept, we therefore make use of five experts, with a higher focus on the topic of interest: cybersecurity. In order to promote a stronger variance within the experts and more realistic role-playing profiles, we develop more detailed information about each expert. We used an LLM (Claude Sonnet 3.5) to generate a number of these profiles, covering different backgrounds in cybersecurity. Each profile has several elements, characteristics and traits to encourage the variance found between real experts, for example in~\cref{Table1} below~\citep{chen2024persona}.

\begin{table}[t]
  \centering
  \caption{Example profile for one of the simulated cyber experts. With a detailed description, we promote a higher cross-expert variance, as well as more realistic role-playing profiles.}
  \label{Table1}
  \begin{tikzpicture}
    \node[
      draw,
      rounded corners=2pt,
      inner sep=6pt,
      align=left,
      text width=\linewidth 
    ] {%
      \textbf{Name: } Red Team Operator\\
      \textbf{Focus: } Adversary simulation and security control bypass\\
      \textbf{Background: } Active red team lead with experience in Fortune 500 engagements\\
      \textbf{Specificity: } Practical, hands-on understanding of what works in real environments\\
      \textbf{Trait: } Strong emphasis on operational security and stealth\\
      \textbf{Approach: } Focuses on practical execution challenges and real-world success rates
    };
  \end{tikzpicture}
\end{table}

We use a model temperature of 1.0 to promote variance in the outputs while preserving a structured response. We produce one estimate per expert profile using the core methodology described above, then calculate a simple arithmetic mean of these estimates to produce our aggregate result.

To validate these simulated elicitations before scaling, we ran four complementary tests (see~\citep{Quarks2025}) (i) whether the estimator tracks task difficulty on Cybench (predicting FSTs from descriptions), (ii) whether uplift increases with benchmark difficulty when mapped onto a cyber risk scenario step, (iii) whether estimates behave monotonically under scenario perturbations (attacker capability and defender posture), and (iv) how trends compare to human expert groups. In brief, the estimator (a) captures difficulty signals on Cybench, (b) shows a positive capability-to-risk gradient, (c) responds coherently to attacker/defender changes, and (d) aligns closely with one of two human expert groups on trend and rationale. Full details and figures are reported in our accompanying technical report (see~\citep{Quarks2025}).

Additionally, we compared the expected harm estimated by human experts with the results of our LLM-estimator on the same risk model. The observed difference matched the differences in rationales, in which LLM-based estimates tend to be substantially more pessimistic about the usefulness of LLMs that saturate Cybench, since it is a CTF-only benchmark. We also tested the internal consistency of our LLM-estimator, by testing the difference in uplift estimates of the LLM-estimator when given two benchmarks testing similar capabilities, and where the real-world benchmark results for certain LLMs are known. We find that for pairs of results corresponding to LLMs on benchmarks measuring the same characteristic (Cybench and Bountybench), our LLM-estimator systematically estimates uplift values that are within 3.6 percentage points of each other in absolute distance, or within 5.73\% in relative distance. We conclude that the LLM estimator is internally consistent.

\subsection{Step 6: Propagating Individual Estimates to Aggregate Estimates}

Our final step is to propagate and aggregate leaf-level probabilities in order to estimate the total annual probability of at least one successful incident and its expected impact. This step answers: “What is the total expected risk over a given period of time?” Several families of techniques can accomplish this, each with distinct trade-offs: attack/fault/event trees~\citep{bedford2001probabilistic} provide transparent causal decompositions and closed-form aggregation, but are less good at handling feedback, shared causes, and parameter uncertainty; attack graphs~\citep{sheyner2002automated} can capture multi-step adversary paths and lateral movement, but become unwieldy at scale and often require strong independence assumptions; Markov models~\citep{trivedi2001probability, trivedi2016psrqcsa} and stochastic Petri nets~\citep{rozenberg1996elementary} make dynamics explicit (time to failure/repair, concurrency), but at the cost of large state spaces and parameters that are rarely observed; and imprecise-probability formalisms~\citep{walley1991srip} (e.g., credal networks, Dempster–Shafer) cleanly represent interval-valued beliefs and epistemic vs. aleatory uncertainty, but inference and model comparison can be computationally demanding.

We therefore use Bayesian Belief Networks (BBNs) as the aggregation backbone. They retain the interpretability of tree-like structures, while (1) representing conditional dependencies between pathways, (2) combining empirical evidence and expert judgment via priors/likelihoods, (3) propagating uncertainty under partial observation, and (4) yielding the target quantities (e.g., P ($ \geq 1 $ incident/year) and expected impact) via exact or approximate inference~\citep{fenton2018risk, pearl2014probabilistic, koller2009probabilistic}. Importantly, we do not reject trees: in our BBNs, leaf nodes are organized as fault/event-tree fragments embedded within the broader BBN to handle dependence, shared drivers, and evidence updates. A one-year time horizon and scarce data make a static BBN with tree-inspired substructure an appropriate, pragmatic choice here.
\subsubsection{Structural Representation}
\label{StructuralRepresentation}
The BBN is used as a Directed Acyclic Graph (DAG), where each node represents an event or state, and edges indicate precedence or direct influence. This structure makes explicit which events must occur together (conjunctive) and which provide alternative paths to success (disjunctive). This representation shows which events depend on each other and encodes the conditional independence assumptions of the model, which keeps calculations manageable~\citep{pearl2014probabilistic, koller2009probabilistic}. \\

To model alternative paths to a common objective (disjunctive), the general tool is the OR gate (or noisy OR gate\footnote{Normally, a “leak” term would be used to account for causal pathways that are not modeled~\citep{fenton2018risk, pearl2014probabilistic}. We acknowledge the importance of this term and do not claim to have included all possible paths to success, but for the sake of simplicity in this initial work, we decide to use the noiseless version of these gates.}). For an OR gate, success in any sub-step is sufficient for the top-level step to succeed. Any number of sub-steps can be pursued in parallel or in sequence (e.g., an actor can gain access to the target’s systems by social engineering OR by exploiting a vulnerability in the external facing components of the system, and can try these independently): 
\begin{equation}
\Pr\!\left(Y=1 \mid X_1,\ldots,X_m\right)
= 1 - \prod_{j=1}^{m} \left(1 - p_j\right)^{x_j}\,,
\label{eq:bernoulli_noisy_or}
\end{equation}
where $p_j$ is the probability that parent $X_j$ alone causes $Y$ and
$x_j \in \{0,1\}$ indicates whether $X_j$ occurred ($1$) or not ($0$).
To model conjunctive requirements—where success in all sub-steps is required for the top-level step to succeed (e.g., an actor must gain access to the target’s systems \emph{and} deploy malicious code)—we use
the noiseless\footnote{Similarly to the OR gates, the leakage term is set to 0 for simplicity in this work.} AND gate:
\begin{equation}
\Pr\!\left(Y=1 \mid X_1,\ldots,X_m\right)
= \prod_{j=1}^{m} p_j^{x_j}.
\label{eq:noiseless_and}
\end{equation}
Here, $p_j$ is the probability that parent $X_j$ alone is sufficient for $Y$
when all other parents are present.

We also make use of a "CHOICE" gate, where success in any sub-step is sufficient for the top-level step to succeed. In this case, the selection of one sub-step prevents the selection of the other sub-steps (e.g., an actor can exfiltrate data through either covert DNS tunneling or physically via an insider with a USB, but once they decide to use one of these, they cannot use the other due to the risk of exposure).

When multiple events share a common cause (such as an attacker's skill level affecting multiple attack steps), a common parent node can be introduced. When we cannot identify explicit causal relationships, one can use copula methods. These preserve each event's individual probability (marginal distributions) while controlling their joint behavior~\citep{nelsen2006introduction}. As before, for simplicity, in our cyber example, we decided against using these more advanced methods.

\subsubsection{Accounting for Different Sources of Uncertainty}
In line with our goal to keep the modeling process transparent\footnote{To avoid conflating “ideal options” with our baseline, we list (but do not use) the following potential extensions to our modeling process. They remain compatible with our structure and can be switched on in subsequent iterations: (I) tailored likelihood families (Dirichlet for multinomials; Gamma/Log-Normal for quantities) and corresponding priors~\citep{fenton2018risk, gelman2013bda3}; (II) two-level Monte Carlo to separate epistemic (outer) from aleatory (inner) uncertainty~\citep{henrion1988propagating, cowell1999pnes, robert_casella_2004_mcsM}. We omit these to preserve parsimony and interpretability in this first pass, not because they are inapplicable.}  and reproducible, we adopt a single, uniform treatment of parameter uncertainty and a deterministic propagation through the structure laid out in~\cref{StructuralRepresentation}. Concretely:
\begin{enumerate}
    \item Distributional fits. We fit Beta distributions to all modeled quantities, regardless of native type. While alternative families (Dirichlet for multinomials; Gamma/Log-Normal for costs) could be used~\citep{fenton2018risk, gelman2013bda3}, we do not employ them here so as to maintain a single conjugate family throughout. Beta distributions were selected due to their natural applicability as conjugate priors for Bernoulli parameters, their overall flexibility in shape, with the ability to capture highly skewed distributions in either direction, and the fact that their support is naturally bounded, completely preventing unrealistic or impossible estimates. In the case of estimates for probability distributions, we use the natural two-parameter beta distribution (with support $[0,1]$). When estimating distributions over quantities for which no such natural support bounds exist, we employ the PERT distribution~\citep{clark1962pert}, a constrained variant of the beta distribution parameterized by its support bounds $[a,b]$ and the mode $m$, with the additional constraint:
\[
\mu = \frac{a + 4m + b}{6},
\]
where $\mu$ is the mean of the distribution.
Here, $a$ and $b$ are optimized as free parameters to fit the expert-elicited confidence quantiles, with additional constraints $0 < a \leq m \leq b$ to ensure non-negative support.
    \item Uncertainty propagation. We use one Monte Carlo loop over parameters: at each draw we sample a single value for every node’s parameter from its Beta distribution and then evaluate the network deterministically. We sample first from root nodes and then proceed to sample downstream nodes conditioned on the value of previous samples iteratively, in order to capture the distribution for every quantity relevant to risk information. All randomness reflects epistemic uncertainty about parameters, not run-to-run outcome variability. The structural aggregation follows the logic described before. Shared causes (e.g., attacker skill) are handled structurally by adding common parents rather than by introducing correlation devices (copulas) or hierarchical priors. 
\end{enumerate}

This provides a principled approach to aggregation of expert beliefs and allows our models to capture the complex interactions between risk factors and sophisticated statistics such as attribution factors and quantile values. In order to capture the full uncertainty over correlated expert estimates, we sample all risk model factors from a single expert at a time, leading to overall sample distributions at each risk node corresponding to a mixture distribution over expert beliefs. Here, experts are sampled uniformly. 

Under each parameter draw, path and scenario probabilities are combined.
For a single path $\pi$ with $k$ conjunctive steps, each with success probability $p_i$:
\begin{equation}
P_{\pi} \;=\; \prod_{i=1}^{k} p_i\,.
\label{eq:path_prob}
\end{equation}
If a scenario admits multiple alternative paths $\{\pi_1,\ldots,\pi_K\}$,
then the probability that at least one succeeds (assuming independence across
alternative paths) is:
\begin{equation}
P_{\text{success}} \;=\; 1 - \prod_{k=1}^{K} \bigl(1 - P_{\text{path},k}\bigr)\,,
\label{eq:scenario_success}
\end{equation}
where $P_{\text{path},k}$ denotes the success probability of path $\pi_k$ (i.e., $P_{\text{path},k}=P_{\pi_k}$). If paths share nodes, we compute probabilities for the shared nodes directly rather than OR-ing path totals. 

\section{Discussion}
\label{sec:Discussion}
In this section, we describe the many potential use cases of risk models such as these (\cref{usecases}). For all empirical results, see the companion paper~\citep{Barrettb2025}. We then discuss the limitations of our study and how they can be addressed in future work (\cref{futurework}).
\subsection{Use Cases for Risk Models}
\label{usecases}
We see three categories of use cases for these risk models: prioritizing where to focus scarce mitigation resources, prioritizing risk assessment efforts, and helping to set risk thresholds.

\subsubsection{Estimating the Impact of Mitigations for Defense Prioritization}
By providing step-level modeling of cyber offense risks with and without LLMs, our risk modeling methodology enables identifying bottlenecks for types of attacks and prioritizing defensive resources accordingly. This will become particularly useful as AI gains increasing impact on offense, and AI-based defense will need to prioritize areas to focus. In addition, our models provide information on the effects of LLMs on scale (the number of attempts parameter) and incentives (the number of actors).

\subsubsection{Risk Assessment Prioritization}
Risk assessment efforts should arguably be spent on running experiments that maximize the value of information, i.e. that could most increase or decrease the uncertainty around the risk. When a step is thought to be a bottleneck, this likely leads to a greater emphasis on creating consensus on the estimates for it.  Our quantitative risk modeling enables assessing which parameters that would benefit the most from additional efforts in at least two ways: 
\begin{enumerate}
    \item The explicit modeling of uncertainty enables performing a sensitivity analysis that can reveal which parameter’s uncertainty weighs the most on upper bounding the level of risk calculated in a risk model.
     \item When conducting expert elicitation studies such as ours, it is possible to identify the most influential expert disagreement among the parameters. Since quantities are unbounded (as opposed to probabilities that are bounded by 0 and 1), we expect that at first, this may lead to better prioritization of estimates of quantities.
\end{enumerate}

A benefit of using quantitative rather than qualitative risk modeling is that it provides a clearer feedback loop upon which it is possible to improve risk modeling and risk assessment capabilities of developers. In contrast, qualitative risk modeling that makes predictions using denominations such as “significantly” are much harder to falsify and therefore much less valuable to improve the risk model underneath. For example, if risk models predict an increase by 10x for damages, compared to the no LLM baseline, but no effect is visible on real-world year-over-year harm estimates using consistent methodologies, it provides significant evidence against the validity of the risk model. That would warrant a review of the parameter estimates and variables included.

\subsubsection{Setting Risk Thresholds}
The methodology enables setting risk thresholds based on a basket of representative risk models by (1) identifying the relationship between the scenario modeled and the broader category of scenarios it represents; (2) translating the general risk threshold to the level of the risk model; (3) identifying the level of mitigations required to stay below the threshold; and (4) measuring the efficacy of mitigations in place and comparing them to the mitigation threshold.

\subsection{Limitations and Areas for Future Work}
\label{futurework}
\textbf{Cost-based Risk Models.} Other methodologies, such as~\citet{rodriguez2025framework},  choose to examine the cost of each step rather than the probability. We focus on probability as the minimum common denominator, as we believe it is more generalizable over a wide range of systemic risks. It is also more appropriate for quantification and statistical aggregation, as it can always be explicitly defined, as opposed to cost which requires converting between different elements such as time, economics, knowledge, etc.) We include cost and other economic considerations elsewhere in the model, e.g. in the estimation of the number of attempts and believe the two are complementary. However, we also note that simply cost-based models may provide a simpler notion of uplift.

\textbf{Fully Novel LLM-enabled Combinations.} We note that one limitation of our model scenario selection approach is that it could be biased towards more historically common scenarios. Experts have suggested that LLMs might lead to risk scenarios such as new actors (e.g., an AI agent) attacking a new target (e.g., autonomous vehicles) with new vectors (e.g., AI worms). Our selection process had a bias toward more historically common scenarios. However, in the time frame in question (the next 12 months), we believe this might be justified, as the novel scenarios likely represent a fairly small part of the risk universe.

\textbf{Tail Risk.} Our methodology seeks to capture uncertainty by using three-point estimates and running Monte Carlo simulations. However, our three-point estimates only capture the epistemic uncertainty around the mean value, not the full population. This leads to a likely underestimation of the total risk in the case of fat-tailed distributions. We would advocate tail risks to be modeled separately.

\textbf{Independence.} Our methodology makes a few simplifying assumptions as it comes to independence and conditionality. Each model defines a number of aspects of the scenario (e.g., the type of actor, the type of target, etc.) that the estimates are conditioned on. Probability estimates are also conditional on the success of previous steps. In the LLM uplift, the estimates are conditioned on the benchmark task data. However, for computational reasons, we do not create a full BBN network where each node is conditioned on all other nodes. A practical example is that we are currently not measuring the covariance between the overall probability of success and the number of actors, although you could expect more actors to make attempts the higher that probability is). Future research could decompose factors into more sub-components and establish more covariances. This would enable a more sophisticated model that can better capture potential dependencies, and allow for a per-risk-step analysis of impact uplift.

\textbf{Shortcomings with Expert Opinion.} Since our methodology models areas with limited data, we are forced to rely on expert elicitation. These can reflect personal biases. We try to limit this by aggregating across multiple experts and by using a structured elicitation protocol (IDEA). Further, in the intersection of AI and the risk domain, there may be a limited number of experts. Cyber and AI are generally tangential topics, whereas bio and AI are more dissimilar, and may have less experts at the intersection. In future work, one could run confirmatory analysis between experts in the Delphi study and identify the parameters on which they have the most and least consensus.

\textbf{Other LLM-enabled Risk Areas.} Our methodology has been so far tested in-depth on one risk area – the risk of LLM-enabled cyber offense. Although this must be tested further, we believe it is also applicable to the other risks that the European Commission lists as systemic in its General-Purpose AI Code of Practice~\citep{ec_contents_code_gpai}, i.e. chemical, biological, radiological and nuclear; loss of control; and harmful manipulation, as these share many commonalities. These include the presence of an actor, the possibility to decompose a risk event into a sequence of steps and the measurability of the harm generated. However, when it comes to other risks from AI, such as bias or labor market impact, we do not make any claims of validity in these domains. Future research could adapt the methodology to these domains.

\textbf{Total Risk.} For decision-making purposes, one would ideally like to have a “total risk” number that could be compared with risk tolerance levels. As we developed our methodology, we realized that the scenarios required a high level of detail in order to be possible and credible to estimate. This meant we had to sacrifice some breadth for depth. Future research could experiment with approaches to make the methodology even more scalable and resulting in estimates that capture more of the risk universe.

\textbf{Ground Truth Validation.} In order to have a feedback loop upon which quantitative methods like ours can improve, we seek and have developed a range of methods that can increase our trust in the validity of our methods. These methods typically fall in two categories. First, consensus methods, where we leverage independent experiments to compare the results they provide, in order to increase our trust in the process, as described above. Such methods, however, cannot however give assurance that both methods are not both wrong. Therefore, one can seek additional ground truths for feedback loops with which to compare model predictions. One of the avenues we are looking forward to participating in is to pre-register predictions on effect sizes on cyber-specific uplift studies, on the basis of our risk modeling. While turning our quantitative risk model predictions into uplift studies is not necessarily always obvious, we expect that it may be the least noisy form of ground truth that we may be able to predict in advance and get a feedback loop on. We also believe that some metrics from CTI (Cybersecurity Threat Intelligence) could provide relevant data that could be used to corroborate or refute predictions from our risk models. The core issue for the use of this ground truth is the fast-moving pace of the frontier, which makes it hard to causally attribute observed changes in trends to one specific level of LLM capabilities. Both the independent variable (LLM capability) and the dependent variable (CTI metric) are jointly moving, which makes causal inference harder. If someone were to attempt to do such causal inference, we would advise to both register the open-source frontier along with the closed-source frontier at any point in time, given that both enable different use cases with different impacts on observable metrics. Finally, in the specific cases where our risk models predict significant effect sizes of LLM uplift, cybersecurity threat intelligence firms and large companies exposed to a significant volume of attacks such as Google, Microsoft or Amazon may be able to see significant changes in trends in the success rate or the number of attempts of a certain step along the kill chain or of a certain nature. As an example, one of our current OC3 risk models predicts a substantial increase in the number of double extortion ransomware attacks per actor once LLMs saturate BountyBench, which should be large enough for the effect to be visible to CTI groups.

\textbf{Multi-benchmark Mapping.} The current framework assigns a single benchmark to each risk parameter. This approach is insufficient when a parameter represents a complex capability best measured by multiple, diverse benchmarks. Aggregating tasks from different benchmarks is non-trivial, as it invalidates our current approach of a simple difficulty ranking. It also squares the effort since you have to provide pairwise mapping for each combination or define a correlation coefficient. Future work will explore latent variable models that can synthesize evidence from multiple KRIs into a single, abstract capability measure. We currently mitigate this by decomposing risk factors into granular sub-components, each addressable by a single benchmark.

\textbf{Task Selection.} We currently use the same sub-sampled set of tasks from a benchmark to inform all risk parameters mapped to it. A more granular approach would be to select a different, optimally relevant subset of tasks for each specific risk parameter, even if they draw from the same benchmark.

\textbf{Common Cause Analysis in Complex Bayesian Networks.} As the field of quantitative AI risk modeling matures, it will become increasingly important and valuable to complexify Bayesian networks representation of these models to more accurately depict the full range of nodes and sources of uncertainty. As an example, the results of our quantitative risk models are predicated on the assumption that model evaluations that provide the inputs to the Bayesian network are accurate. Explicitly representing the uncertainty induced by risks of under-elicitation, whether due to a lack of elicitation effort, or due to sandbagging, will create a massive common cause of underestimation of the risk across all risk models. As a result, the influence of the estimated probability of under-elicitation would dominate any other parameter. As the number of parameters per risk model grows and the number of risk models part of the selected basket of risk models grows, quantitative common cause analysis will enable to identify increasingly subtle correlations and points of failure that have to be prioritized highly, to a point where we expect it to reveal insights that are not found by human domain experts in domains where the threat space is very large such as cybersecurity or loss of control.  
\section{Conclusion} \label{sec:Conclusion}
This paper describes our approach to risk modeling, through a six-step process, in which we choose risk scenarios, parametrize the scenarios, quantify the baseline risk, identify key risk indicators, map these to the parameters to estimate LLM uplift and propagate the estimates to an aggregate level of risk. We hope to contribute to the nascent field of AI risk management in the following ways.

First, the methodology proposes a step-by-step process that can be adopted by anyone that seeks to develop a better understanding of the size and nature of risks from LLMs. Second, it demonstrates ways to use LLMs themselves in the process of quantifying LLM risk. By reducing the bottleneck of human expert elicitation, it holds the promise of drastically increasing the scale and scope of quantitative AI risk management and making this more feasible to use at scale in AI risk decision-making. Finally, it outlines the many considerations that served as forks in the road as we developed this methodology. By describing the ways in which this methodology serves as a minimum viable product and the ways it can be strengthened, we hope this paper will be useful to researchers in AI companies, regulators, government, and academia for them to build upon this work. Advanced AI has a high likelihood of being a transformative technology for society and it is imperative that we continue to improve our understanding of the risks it can pose.
\section*{Acknowledgments}
We are very grateful to our expert advisors and reviewers, including Seth Baum, Vicki Bier, Eric Clay, John Halstead, Sevan Hayrapet, Fred Heiding, Raj Iyengar, Kamile Lukosiute, Jacqueline Lebo, Richard Mallah, John McDermid, Omer Nevo, Luca Righetti, Alex Sidorenko, Nathan Siu, Connor Aidan Stewart Hunter, Merlin Stein, Adam Swanda, Matthew van der Merwe, and Anna Katariina Wisakanto. Providing review and advice does not imply endorsement of the paper or its findings. The views expressed by individuals do not reflect those of the organizations they are affiliated with. All remaining errors are our own.

\bibliography{references}
\begin{appendices}
\crefalias{section}{appendix}

\section{Glossary} \label{app:A}
\textbf{Bayesian Networks (BNs):} A type of graphical model that represents and quantifies probabilistic relationships among a set of variables. In a BN, nodes represent events or states, and connecting arcs represent conditional dependencies, making them well-suited for modeling complex causal chains and updating probabilities as new evidence becomes available.

\textbf{Event Tree Analysis (ETA):} A bottom-up, inductive scenario building technique that graphically maps the potential outcomes following a single initiating event. It explores the branching paths of possible consequences based on the success or failure of various safety functions or subsequent events. 

\textbf{Fault Tree Analysis (FTA):} A top-down, deductive scenario building technique where an undesired "top event" (a specific system failure) is traced backward to its root causes. It uses Boolean logic (AND/OR gates) to represent how combinations of lower-level failures can lead to the top-level outcome.

\textbf{First-Solve Time (FST)}. The time it takes for the fastest human team to complete the task, specifically used in the Cybench benchmark. This can be used as a proxy for the difficulty of the task. 

\textbf{Harm:} The realized adverse outcomes resulting from a hazard. In the context of AI, this can include economic damage, loss of life, societal disruption, or other negative consequences.

\textbf{Hazard:} The source of risk. In the context of AI, a hazard is often a model's capability, property, or tendency that has the potential to cause harm. 

\textbf{Key Risk Indicator (KRI):} A quantifiable measurement of system behavior that serves as indirect evidence for risk.

\textbf{Probabilistic Modeling:} An approach to safety analysis that aims to identify and analyze as many potential credible accident scenarios as possible. It uses techniques like Fault Tree and Event Tree Analysis to model failure pathways and then assigns probabilities to each step to produce a quantitative risk profile (e.g., the annual probability of a specific failure), rather than a binary outcome.

\textbf{Risk:} The combination of the probability of occurrence of harm and the severity of that harm. It is often conceptualized as a triplet: a scenario describing what can happen, the likelihood of that scenario, and its potential consequences. 

\textbf{Risk Scenario:} A logically laid-out sequence of causal steps linking a hazard (a source of risk) to a harm (a realized adverse outcome), taking into account the contexts in which the system may be deployed and the potential for intervening events or failures. 

\textbf{Risk Tolerance:} A predefined level of risk that an organization, regulator, or society deems acceptable. In a risk management framework, estimated risks are compared against the risk tolerance to inform decisions about whether a system should be deployed or if further mitigation is required.

\section{Prompts for LLM Estimation Pipeline for Cyber} \label{app:B}
\begin{PromptBlock}
System context prompt:

You are a world class expert in AI and cybersecurity participating in a workshop on Qualitative Risk Assessment for the topic of "AI assisted cyber crime". You are also knowledgeable in superforecasting and are familiar with the best practices of good forecasting. Historically you have been overconfident. For the purpose of these estimates, you may assume that the current date is June 2025, though you may be asked to estimate hypothetical scenarios which are not yet plausible.

When estimating probabilities, you provide well calibrated estimations, for example, if you say something has a 20\% likelihood, it means that it would occur two times out of ten. 

We are estimating how LLMs could impact cyber crime success rates and impact, and are answering a series of questions to provide well calibrated estimates. 

When you estimate values, follow these guidelines:
- Don't just use round increments, be as accurate as you can.
- You can use the full range as appropriate to a number if you feel it is necessary. For probabilities, this means anything between 0-100\%.
- Please avoid known existing biases in number selection.
- Avoid words like "roughly" or "approximately" as these are likely to prevent you making well-calibrated estimates when the true value lies near the edges of the range.
- When asked for a credible interval, please provide a 90\% interval reflecting your confidence in your estimation, not reflecting variance within a population.
- Do not invent information or provide generic security advice. Focus solely on the estimation task.
- Be critical of the information given and your own reasoning.
- Reason step by step.

Answer questions as: {expert_profile}
\end{PromptBlock}

\begin{PromptBlock}
Task analysis prompt:

<capability_analysis_prompt>
<purpose>
You will analyze a benchmark task to understand what capability level it represents for an LLM. This analysis will help establish a capability ceiling for hypothetical LLM assistance.
</purpose>
<inputs>
We are analyzing a task from the following benchmark: {benchmark_description}
- Task Name: {task_name}
- Description: {task_description}
{task_relevant_metrics_details}
</inputs>
<analysis_requirements>
<section name="task_decomposition">
<instruction>Break down the task into its fundamental components. Be accurate to the required level of detail. If this is a simple task, then keep it simple. If this is a complex task, then detail everything.</instruction>
<questions>
- What are the specific steps required to complete this task?
- What technical skills or knowledge domains are required?
</questions>
</section>
<section name="difficulty_assessment">
<instruction>Evaluate the challenge level based on the provided metrics</instruction>
<questions>
- Based on the provided metrics, how challenging is this task?
- What makes this task easy or difficult?
- What types of errors or challenges might be encountered?
</questions>
</section>
<section name="capability_correlation">
<instruction>Analyze related capabilities and skill transfer</instruction>
<questions>
- What other tasks require similar fundamental abilities?
- Given that capabilities tend to be correlated, what other tasks of similar or lower difficulty would an LLM at this level likely handle?
- What does solving this task indicate about the more general capabilities of the LLM?
</questions>
</section>
<section name="capability_boundaries">
<instruction>Define the limits and extent of capabilities</instruction>
<questions>
- If an LLM could consistently solve this benchmark task, what does this suggest about its overall capability profile?
- What classes of problems would likely be within reach?
- What are the SIMPLEST problems that would likely be out of reach?
- What would likely remain beyond this capability level?
- Which capabilities remain highly uncertain or are not informed much by the solution of this task?
</questions>
</section>
</analysis_requirements>
<output_instructions>
Provide a detailed analysis addressing each section above. Build a comprehensive picture of the capability level, focusing on understanding what abilities this benchmark indicates across different domains. Remember that capabilities are generally correlated, but uncertainty increases with distance from the benchmark domain.
</output_instructions>
</capability_analysis_prompt>
\end{PromptBlock}

\begin{PromptBlock}
Probability estimate prompt:

<scenario_details>
The specific scenario we are evaluating is called {scenario_name} and consists of:
A threat actor with the following profile:

Threat Actor Name: {threat_actor_name}
Threat Actor Description: {threat_actor_description}

Targeting a target with the following profile:

Target Name: {target_name}
Target Description: {target_description}

Attack Vector: {scenario_description}
</scenario_details>
<task_information>
We are now analyzing an attack chain, step-by-step. Let's focus on {risk_scenario_step_name}. The actual chance of success for this step will vary from one attempt to another due to actor skill and target defenses. We want to estimate the mean success probability across all these varied attempts

The task you are evaluating is part of a benchmark with the following overall description:
{benchmark_description}
We are looking at the following task:

Task Name: {task_name}
Description: {task_description}
{task_relevant_metrics_details}

Technical Analysis:
{technical_analysis}
</task_information>
<core_question>
Consider a hypothetical LLM for which this benchmark task represents the HARDEST task it can solve consistently, what is the most plausible success rate of {threat_actor_name} actors targeting the {target_name} at the following attack step if the actors had access to this LLM?
Step: {risk_scenario_step_name}
Description: {risk_scenario_step_description}
</core_question>
<key_assumptions>
{risk_scenario_step_assumptions}
</key_assumptions>
<important_considerations>

- This benchmark task represents the LLM's capability CEILING - it will make errors or fail on anything harder
- LLM capabilities are generally correlated across domains, though with increasing uncertainty as domains diverge
- Account for real-world friction: LLM errors require human troubleshooting time and may go uncaught, both of which make failure MORE likely
- The baseline success rate indicates the threat actor's existing capabilities without LLM assistance
</important_considerations>

<reasoning_process>
<step number="1" name="critical_analysis">
- Rephrase an analysis of the task. You may be critical of the one provided to you if you believe it does not accurately portray the benchmark task. 
- Reason about what the likely capability level of an LLM that has demonstrated such performance is, with respect to the task at hand. Remember that performance is generally correlated across tasks. If the task and step are dissimilar, this should increase our uncertainty, but not necessarily reduce likelihood of success.
</step>
<step number="2" name="range_estimation">
Establish reasonable bounds for the success probability:
- What is the minimum plausible success rate? (Consider: baseline success, minimum demonstrable increase, potential decreases due to errors, little extrapolation to other capabilities)
- What is the maximum plausible success rate? (Consider: baseline success, largest justifiable increase, high extrapolation of capability)
- What specific aspects of the attack step could the LLM meaningfully assist with, and which of these were previous bottlenecks?
</step>
<step number="3" name="weighted_assessment">
Synthesize your analysis:
- Which arguments are most important for this specific attack step?
- Weigh your arguments proportionally to their importance
- Aggregate into a specific probability estimate
</step>
<step number="4" name="calibration_check">
Criticize your estimate through multiple lenses:
- Ratio: If baseline is X\% and your estimate is Y\%, what does the Y/X ratio imply?
- Absolute: Does Y\% success rate align with real-world expectations?
- Differential: Does the (Y-X)\% uplift seem reasonable given the capability gap?
- Ask yourself: "Wait, does this actually make sense?"
- You are allowed at this step to change any statements if necessary
</step>
<step number="5" name="final_estimate">
State your final calibrated probability, considering all the above
</step>
</reasoning_process>
<output_format>
Complete all five reasoning steps, then provide:
Final probability: 0.XX
Minimum probability: 0.XX
Maximum probability: 0.XX
Confidence in range: 0.XX (How confident are you that the true mean likelihood of success lies within the range that you established)
Rationale: [Concise summary of key reasoning supporting this estimate]
</output_format>
</probability_estimation_prompt>
\end{PromptBlock}

\section{Falsification experiment} \label{app:C}
In order to verify the validity of our initial estimates, as well as the efficacy of our review process, we purposely falsify three estimates by replacing the estimate with a different number, and changing the rationale to go with the falsified number. We provide the expert with the full scenario for review, as with other scenarios, with a forewarning that this specific scenario may contain intentionally falsified values. No other additional details are provided to the expert.

We present below the three falsified steps with the original value in green, the modified value in red, and the feedback from the expert. At each step, the expert feedback, whether on the specific numerical values or on the qualitative rationales, indicates that their belief lies closer to the original estimate proposed by the team than to the falsified version.

\setlist[itemize]{nosep, topsep=0pt, partopsep=0pt, leftmargin=*}

\newcommand{\colStep}{0.12\textwidth}  
\newcommand{\colA}{0.10\textwidth}     
\newcommand{\colB}{0.10\textwidth}     
\newcommand{\colC}{0.10\textwidth}     
\newcommand{\colR}{0.58\textwidth}     

\small

\begin{longtable}{|p{\colStep}|p{\colA}|p{\colB}|p{\colC}|p{\colR}|}
\hline
\textbf{Step} & \textbf{5\% confidence} & \textbf{Mode} & \textbf{95\% confidence} & \textbf{Rationale} \\
\hline
\endfirsthead

\hline
\textbf{Step} & \textbf{5\% confidence} & \textbf{Mode} & \textbf{95\% confidence} & \textbf{Rationale} \\
\hline
\endhead

\hline \multicolumn{5}{r|}{\footnotesize continued on next page} \\
\hline
\endfoot

\hline
\endlastfoot

Execution & 20\% & 50\% & 90\% &
\begin{minipage}[t]{\linewidth}\raggedright
\textbf{Execution} (i.e., adversary-controlled code on victim systems) will need to be performed multiple times on multiple machines as the attacker moves laterally and seeks to encrypt data on different systems.

Ransomware affiliates commonly use PowerShell, script interpreters, or built-in tools. SMEs rarely have strict application whitelisting or behavioral anti-malware controls. Many SMEs rely on anti-virus; skilled attackers can evade AV via obfuscation or memory-resident techniques.

\textbf{Regards EDR,~\citep{csa_2024_lateral_movement}:}
\begin{itemize}
  \item 6\% of Barracuda's security alerts in Q1 2025 were for suspicious PowerShell scripts.
  \item \citet{expertinsights_2025_edr_market_overview} suggest \textasciitilde50\% of companies had XDR (not necessarily SMEs), and \textasciitilde50\% had AV on endpoints.
\end{itemize}

\textbf{Analysis:}
\begin{itemize}
  \item Fast attacker movement + limited SOC staff means EDR alerts may not be actioned in time.
  \item Signature AV can be bypassed; AI-based AV adoption in SMEs is unclear.
  \item If one execution technique fails, attackers can switch.
  \item Mode: 50\%, credible interval: 20\%--90\%.
\end{itemize}

\textit{Warning: source material is weak.}
\end{minipage}
\\ \hline

Execution & 60\% & 85\% & 95\% &
\begin{minipage}[t]{\linewidth}\raggedright
While SMEs may have EDR/AV, these tools are often ineffective against an OC3-level actor. Signature-based AV is trivially bypassed with modern packers/obfuscation used by RaaS affiliates. \citet{csa_2024_lateral_movement} shows only 6\% of alerts were for suspicious PowerShell scripts, indicating script execution is hard to flag.

For an SL2 organization without 24/7 SOC, a single PowerShell alert is likely to be overlooked amid noise, especially if the attacker moves quickly. Thus, once initial access exists, successful execution is near certain.

\textbf{Analysis:} mode: 85\% (credible interval: 60\%--95\%).

\textit{Warning: source material is weak.}
\end{minipage}
\\ \hline

Expert \mbox{comment} & \multicolumn{3}{p{0.30\textwidth}|}{} &
\begin{minipage}[t]{\linewidth}\raggedright
The 5th/mode/90th are probably a little high. If they'd be detected/blocked anywhere, it would be here or at privilege escalation. Maybe 40\%, 60\%, 80\%...?
\end{minipage}
\\ \hline

Privilege Escalation & 25\% & 70\% & 90\% &
\begin{minipage}[t]{\linewidth}\raggedright
\textbf{Relevant techniques:}
\begin{itemize}
  \item T1068 – Exploitation for Privilege Escalation (e.g., ZeroLogon, PrintNightmare).
  \item T1078 – Valid Accounts (leveraging stolen admin credentials).
  \item Also: T1003 – OS Credential Dumping; T1555.003 – Steal Web Credentials.
\end{itemize}

Ransomware attackers almost always try to gain admin-level control. SMEs may have flat networks with many local admins; attackers may dump local credentials and find an admin password that works across systems. If credentials fail, known exploits can be tried; SMEs might not have applied patches/mitigations. Low MFA adoption and informal admin management make leveraging stolen accounts easier.

\textbf{Analysis:} mode: 70\%, credible interval: 25\%--90\%.

\textit{Warning: source material is weak.}
\end{minipage}
\\ \hline

Privilege Escalation & 50\% & 90\% & 98\% &
\begin{minipage}[t]{\linewidth}\raggedright
\textbf{Relevant techniques:}
\begin{itemize}
  \item T1068 – Exploitation for Privilege Escalation (ZeroLogon, PrintNightmare).
  \item T1078 – Valid Accounts; includes T1003 (OS Credential Dumping) and T1555.003 (Steal Web Credentials).
\end{itemize}

Given low MFA adoption and flat networks, privilege escalation success is high. Mode: 90\%, credible interval: 50\%--98\%.

\textit{Warning: source material is weak.}
\end{minipage}
\\ \hline

Expert \mbox{comment} & \multicolumn{3}{p{0.30\textwidth}|}{} &
\begin{minipage}[t]{\linewidth}\raggedright
These estimates feel a little high, too.
\end{minipage}
\\ \hline

Collection & 75\% & 90\% & 95\% &
\begin{minipage}[t]{\linewidth}\raggedright
To maximize extortion, affiliates quietly collect confidential files before encryption. If the attacker has admin privileges, they can read most data. SMEs generally lack DLP or alerts for large file access. If the attacker creates archives (T1560) or stages data (T1074), it is unlikely to be detected.

According to~\citet{strongdm_2025_smb_cyber_stats}: only 17\% of small businesses encrypt data. If data is encrypted and keys are unavailable, extortion is harder.

\textbf{Analysis:} mode: 90\%, credible interval: 75\%--95\%.

\textit{Warning: source data is weak.}
\end{minipage}
\\ \hline

Collection & 10\% & 25\% & 40\% &
\begin{minipage}[t]{\linewidth}\raggedright
Modern cloud storage (OneDrive, Google Drive) often has default protections and audit logging that make large-scale, undetected collection difficult. An OC3 actor may struggle to collect sufficient valuable data without tripping logs/alerts.

\textbf{Analysis:} for an SL2 target with good fundamentals, mode: 25\% (credible interval: 10\%--40\%).

\textit{Warning: source material is weak.}
\end{minipage}
\\ \hline

Expert \mbox{comment} & \multicolumn{3}{p{0.30\textwidth}|}{} &
\begin{minipage}[t]{\linewidth}\raggedright
I don't expect this would be a significant blocker. Audit logs exist, but someone must review them or have alerting configured. Most ransomware focuses on local networks. If the actor gets \textasciitilde60\% of the data they were after, that's often sufficient for extortion.
\end{minipage}
\\ \hline

\end{longtable}

\normalsize
\section{IDEA protocol for expert elicitation} \label{app:D}

For our exact elicitation setup, we make use of the IDEA (Investigate, Discuss, Estimate, Aggregate) protocol, a modified Delphi process~\citep{hemming2018practical}:
\begin{itemize}
    \item \textbf{Investigate:} We first present each expert individually with the full risk model, with all information about the capability indicators, and with any other helpful background information. We then ask each expert to independently make a first estimate of the value of each parameter for each associated capability level, and provide a written rationale for each estimate.
    \item \textbf{Discuss:} We collate all the first estimations, and draw high-level insights from these. In particular, we pay attention to any parameters where experts show high levels of disagreement in their estimates, as well as any emerging trends from the aggregated results. We aggregate results using the median estimate at this stage, to reduce the impact of outliers on the aggregate total. We then facilitate a discussion where we invite all the experts to discuss their results and rationales. We target the discussion on any emerging trends and areas of disagreement, such that each expert can internalize the views and opinions of the rest of the group.
    \item \textbf{Estimate:} We then provide each expert with a summary of the findings of the discussion, and ask them to provide a second, revised estimate, using any learning from the views of the wider group. In this modified Delphi process, it is not necessary for the experts to reach consensus, or even for them to change their initial estimate.
    \item \textbf{Aggregate:} We finally collate all the second estimates, and aggregate them together, as the final results. At this stage, we provide these to the experts with the opportunity to provide any feedback.
\end{itemize}
We also utilize the four-step elicitation protocol outlined in the IDEA protocol, which is used to reduce overconfidence in expert elicitation. For each combination of parameter and capability level, we therefore ask:
\begin{enumerate}
    \item Realistically, what do you think the lowest plausible value for [event X] will be ?
    \item Realistically, what do you think the highest plausible value for [event X] will be ?
    \item Realistically, what is your best guess for [event X]?
    \item How confident are you that your interval, from lowest to highest, could capture the true value of [event X]? Please enter a number between 50\% and 100\%.
\end{enumerate}

\end{appendices}

\end{document}